\documentclass[a4paper]{jpconf}
\usepackage{graphicx}
\usepackage{epsfig}
\def\PkkE{P({\bf k}_1,{\bf k}_2,E)}
\newcommand{\beq}{\begin{equation}}
\newcommand{\eeq}{\end{equation}}
\newcommand{\be}{\begin{eqnarray}}
\newcommand{\ee}{\end{eqnarray}}

\def\gsim{\buildrel > \over {_{\sim}}}
\begin{document}
\title{Nuclear Physics with Electroweak Probes}

\author{Omar Benhar}

\address{INFN and Department of Physics, ``Sapienza'' Universit\`a di Roma, I-00185 Roma, Italy}

\ead{omar.benhar@roma1.infn.it}

\begin{abstract}
In recent years, the italian theoretical Nuclear Physics community has played
a leading role in the development of a unified approach, allowing for
a consistent and fully quantitative description of the nuclear response to
electromagnetic and weak probes. In this paper I review the main achievements
in both fields, point out some of the open problems, and outline the most
promising prospects.

\end{abstract}

\section{Introduction}

Over the past decades, electron scattering has provided a
wealth of information on nuclear structure and dynamics.
Form factors and charge distributions have been extracted from
elastic scattering data, while inelastic measurements have allowed
for a systematic study of the dynamic response over a broad range
of momentum and energy transfer. Finally, with the advent of the last
generation of continuous beam accelerators, a number of exclusive processes
have been analyzed with unprecedented precision.

In electron scattering experiments the nucleus is seen
as a target. Studying its interactions with the probe,
whose properties are completely specified, one obtains information on the
unknown features of its internal structure. In most neutrino oscillation
experiments, on the other hand, nuclear interactions are exploited to detect
the beam particles, whose properties are largely unknown.

Using the nucleus as a detector obviously requires that its response
to neutrino interactions be quantitatively under control.
Fulfillment of this prerequisite is in fact critical to
keep the systematic uncertainty associated with the
reconstruction of the neutrino kinematics to an acceptable level 
(see, e.g., Ref. \cite{NUINT07} and References therein).

The nuclear response to neutrino interactions is also
relevant to a number of problems in astrophysics. For example, the knowledge of
the mean free path of low energy neutrinos in nuclear matter, over a wide range
of temperature and density, is needed as an input to carry out computer
simulations of both supernov\ae$\ $explosions and neutron star cooling.

The community of italian theorists working on electron-nucleus scattering has a 
long-standing and well established record of achievements. Recently, a significant effort
has been made to generalize the approaches successfully employed in electron scattering
studies to the case of neutrino scattering. This paper is aimed at providing a
short, and by no means exhaustive, summary of the recent results on both the
electromagnetic and weak nuclear responses.

In Section 2, I will briefly review the ongoing discussion on the role of short range
nucleon-nucleon correlations in inclusive electron-nucleus scattering processes, while
Section 3 is devoted to the latest developments of the studies of two nucleon
emission reactions. The results of a number of studies of the neutrino-nucleus cross
section in the kinematical region relevant to the analysis of neutrino oscillation
experiments are discussed in Section 4. Finally, in Section 5 I state the conclusions and
outline my own view of the future of the field.

\section{Correlation effects and final state interactions in 
$e + A \rightarrow e^\prime + X$ processes}

The results of electron- and hadron-induced nucleon knockout experiments have provided
overwhelming evidence of the inadequacy of the independent particle model to 
describe the full complexity of nuclear dynamics. 

While the peaks corresponding to knockout from shell model orbits can be clearly 
identified in the measured energy spectra, the corresponding strengths turn out to be
consistently and sizably lower than expected, independent of the nuclear mass number.

This discrepancy is mainly due to dynamical correlations induced
by the nucleon-nucleon (NN) force, whose effect is not taken into account in 
the independent particle model. Correlations give rise to scattering processes, 
leading to the virtual
excitation of the participating nucleons to states of energy larger than 
the Fermi energy,
thus depleting the shell model states within the Fermi sea. 

The typical energy scale associated with NN correlations can be estimated considering
a pair of correlated nucleons carrying momenta ${\bf k}_1$ and ${\bf k}_2$ much
larger than the Fermi momentum ($\sim250$ MeV). In the nucleus rest frame, 
where the remaining $A-2$ particles have low momenta, 
${\bf k}_1 \approx -{\bf k}_2 = {\bf k}$. Hence, knockout of a nucleon of large 
momentum ${\bf k}$ leaves the residual system with a
particle in the continuum and requires an energy
\beq
E \approx E_{thr} + \frac{{\bf k}^2}{2m}\ ,
\label{corr:en}
\eeq
much larger than the Fermi energy ($\sim$30 MeV). The above equation, where
$E_{thr}$ denotes the threshold for two-nucleon knockout, suggests that large 
nucleon removal energy and large momentum are strongly correlated.
As a consequence, the spectral function $P({\bf k},E)$, yielding the probability 
of removing a nucleon carrying momentum ${\bf k}$ from the target nucleus 
leaving the residual system with energy $E$, is expected to exhibit tails
extending to large ${\bf k}$ and $E$, well beyond the region corresponding 
to the shell model states.

A direct measurement of the spectral function of $^{12}C$ from the
$(e,e^\prime p)$ cross section at missing momentum and energy up 
to $\sim$ 800 MeV and $\sim 200$ MeV, respectively, has been recently 
carried out  by the JLab E97-006 Collaboration \cite{rohe04}. 
The data from the preliminary analysis appear to be consistent with the 
theoretical predictions of sizable high momentum and high energy 
components \cite{benhar89,ramos89,benhar94}.

The search of clearcut evidence of correlation effects 
in the inclusive electron-nucleus cross section at high momentum 
transfer (for a recent review see Ref. \cite{benhar08}) has been pursued by a number 
of experimental and theoretical studies for over three decades. 

As the space resolution of the electron probe is $\sim |{\bf q}|^{-1}$, 
where ${\bf q}$ is the momentum transfer, at large $|{\bf q}|$
scattering off a nuclear target reduces to the incoherent sum of
elementary scattering processes involving individual nucleons.
This is the basic tenet of the Impulse Approximation (IA).
Under the further assumption that there are no final state interactions (FSI) 
between the struck nucleon and the spectator particles, the inclusive 
cross section can be written in the simnple form
\begin{equation}
\left( \frac{d\sigma}{d\Omega_{e^\prime} d\omega} \right)_{PWIA} = \int d^3k dE
\left( \frac{d\sigma}{d\Omega_{e^\prime} d\omega} \right)_{eN} 
 P({\bf k},E) \ ,
\label{dsigma:PWIA}
\end{equation}
where $(d\sigma/d\Omega_{e^\prime} d\omega )_{eN}$ is the cross section
of the scattering process involving a {\em bound} nucleon {\em moving} with 
momentum ${\bf k}$.

It has long been recognized \cite{czyz63} that in the IA regime
short range NN correlations move strength from the quasi-free peak, 
corresponding to electron energy loss $\omega \sim \omega_{QF} = Q^2/2m$, where 
$Q^2 = |{\bf q}|^2 - \omega^2$ and $m$ is the nucleon mass,
to the tails of the inclusive cross section.
While the large $\omega$ region is dominated by inelastic processes, leading to 
the appearance of hadrons other than 
protons and neutrons, the $y$-scaling analysis \cite{sick80} (to be discussed 
in Section \ref{superscaling}) clearly shows that at 
$\omega \ll \omega_{QF}$ the nuclear cross section is mainly due to quasi-elastic 
scattering off nucleons carrying high momenta. 

However, the results of theoretical studies \cite{benhar91,ciofi94} suggest 
that the interpretation of experimental data at low $\omega$ in terms 
of correlations may be hindered by the occurrence of FSI.

The existence of strong FSI in quasi-elastic  scattering has been
experimentally established by the results of $(e,e^\prime p)$ measurements, 
showing that the flux of outgoing protons is strongly suppressed with respect 
to the predictions obtained neglecting FSI.
The observed attenuation ranges from 20-40 \% in Carbon to 50-70 \% in Gold 
\cite{garino92,o'neill95,abbott98,garrow02,rohe05}.

Being only sensitive to FSI taking place within a distance
$\sim |{\bf q}|^{-1}$ of the electromagnetic vertex, the inclusive cross 
section at high momentum transfer is in general largely unaffected by FSI.
However, the effects of FSI can become appreciable, indeed dominant, in 
the low $\omega$ region, where the cross sections become very small.

In inclusive processes FSI have two effects: i) an energy shift $\Delta$ of 
the cross section, due to the fact that the struck nucleon moves in the 
average potential generated by the spectator particles and ii) a 
redistribution of the strength, leading to the quenching
of the quasielastic peak and an enhancement of the tails, as a consequence
of NN scattering processes coupling the one particle-one hole final state to more
complex n-particle n-hole configurations.

As a result, the inclusive cross section can be written in the convolution 
form \cite{benhar91}
\begin{equation}
\frac{d\sigma}{d\Omega_{e^\prime} d\omega} = \int d\omega^\prime \
\left( \frac{d\sigma}{d\Omega_{e^\prime} d\omega^\prime} \right)_{PWIA} \
 f_{{\bf q}}(\omega - \omega^\prime) \ ,
\end{equation}
where $(d\sigma/d\Omega_{e^\prime} d\omega^\prime)_{PWIA}$, given by 
Eq.(\ref{dsigma:PWIA}), is the cross section in the absence of FSI. 
The folding function, embodying 
FSI effects, is trivially related to the spectral function of particle states. 
It can be obtained within the eikonal approximation, i.e. 
assuming that: i) the struck nucleon moves along a straight trajectory with 
constant speed, and ii) the spectator particles act as fixed scattering 
centers. The resulting $f_{{\bf q}}(\omega)$ can be cast in the form
\begin{equation}
f_{{\bf q}}(\omega) = \sqrt{ T_{{\bf q}} }\ \delta(\omega-\Delta) +
( 1 - \sqrt{ T_{{\bf q}}} ) \ F_{{\bf q}}(\omega-\Delta) \ ,
\label{def:ff}
\end{equation}
where the $\delta$-funcion term accounts for the probability that the 
outgoing nucleon does not interact with the recoiling nucleus and
$T_q$ is the nuclear transperency measured in semi-inclusive nucleon 
knockout experiments. The energy shift $\Delta$, whose typical size 
is $\sim$ 10 MeV, is barely visible on the energy loss scale of
inclusive processes at momentum transfer $\gsim$ 1 GeV.  
On the other hand, the redistribution of the strength induced by the 
finite width of the function $F_{{\bf q}}(\omega)$ may lead to a 
large enhancement of the cross section at $\omega \ll \omega_{QF}$.
\begin{figure}[h]
\begin{center}
\includegraphics[width=20.pc]{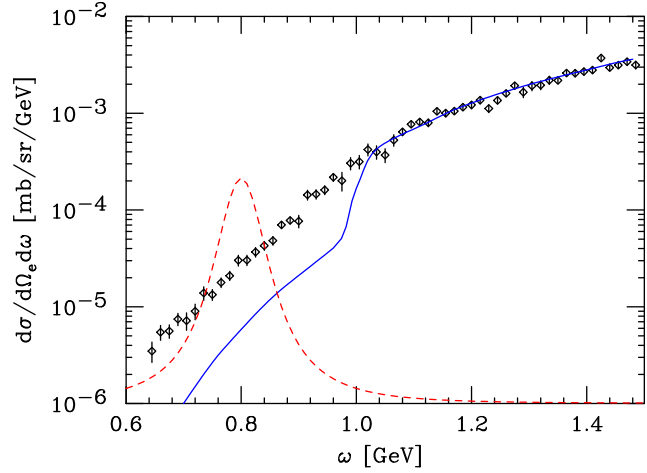}
\end{center}
\caption{Inclusive cross section off uniform nuclear at beam energy 4.0 GeV
and electron scattering angle 30$^\circ$ \cite{day89}, as a function of
the energy loss $\omega$. 
The solid line corresponds to the cross section in the absence of FSI, 
while the dashed line shows the folding function (see Eq.(\ref{def:ff})), 
displayed in linear 
scale and multiplied by a factor 10$^3$ \cite{benhar91}.\label{FSI1} }
\end{figure}
This mechanism is illustrated in Fig. \ref{FSI1}, showing the inclusive 
cross section off uniform nuclear matter, obtained by extrapolation 
of the available data to $A \rightarrow \infty$ \cite{day89}. The solid
line corresponds to the cross section in the absence of FSI, 
while the dashed line shows the folding function (displayed in linear scale 
and multiplied by a factor 10$^3$) \cite{benhar91}. It clearly appears that
going from $\omega \sim$1.2 GeV, where the theoretical cross section 
is in good agreement with the data, to $\sim$ 0.8 GeV, where the 
experiment is severely underestimated, the measured cross section 
drops by more than two orders of magnitude. As a consequence, even a 
a tiny tail of the folding function extending to $|\omega-\omega^\prime|\sim$ 400 Mev
can produce a large enhancement of the cross section at $\omega \sim$ 0.8 GeV

Theoretical calculations show that the FSI effects on the low energy loss 
tail of the cross section, corresponding to $x \gg 1$, where $x=Q^2/2m \omega$ 
is the Bjorken scaling variable, is large, and must be included to reproduce 
the data. 

\begin{figure}[h]
\hspace*{.05in}
\includegraphics[width=19.5pc]{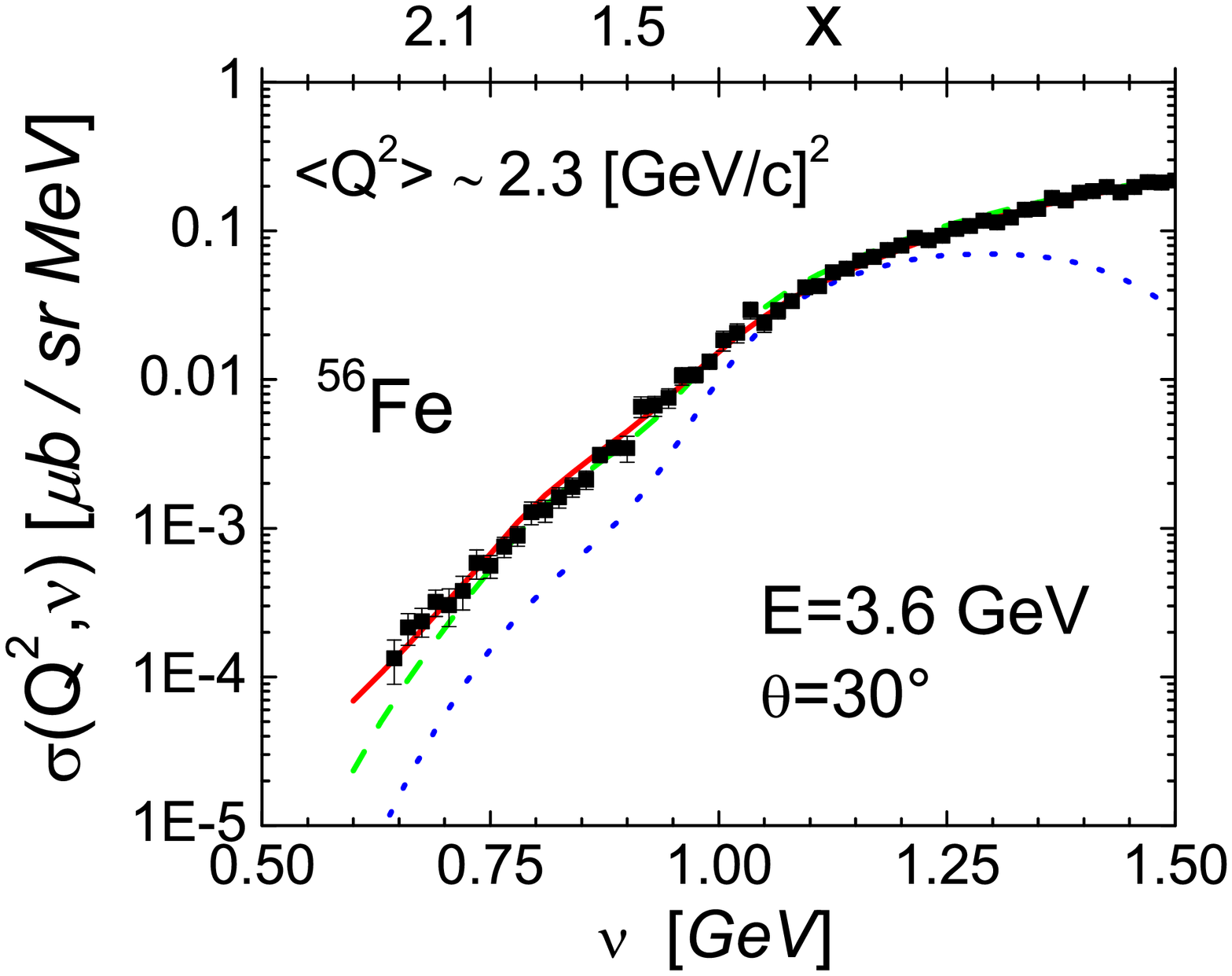}
\includegraphics[width=18pc]{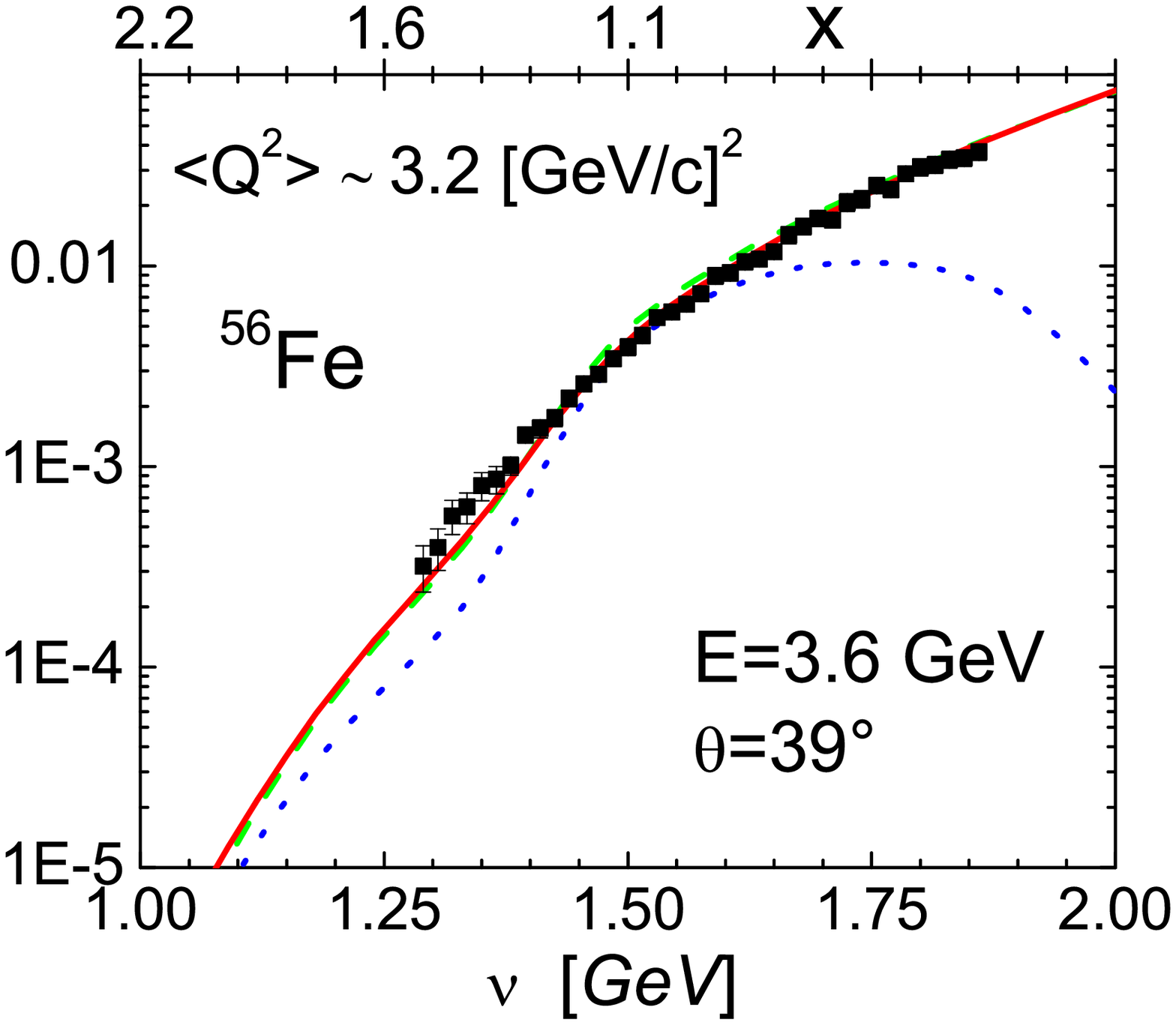}
\caption{Comparison between the measured cross section of the process 
$^{56}Fe(e,e')X$ at beam energy $E = 3.6$ GeV \cite{slac} and the results 
of theoretical calculations including the effects of both short range 
correlations and FSI, represented by the solid lines. The Left and right 
panel correspond to electron scattering angle 30$^\circ$ and 39$^\circ$, 
respectively. The labels of the bottom and top horizontal axes refer to 
the energy transfer 
and the Bjorken scaling variable $x$, respectively. (After Ref. \cite{alvioli}). 
\label{claudio} } \end{figure}

As an example, in Fig. \ref{claudio} the SLAC data corresponding
to iron target, beam energy $E = 3.6$ GeV and electron scattering angle
30$^\circ$ and 39$^\circ$ \cite{slac}, are compared to the results of
Ref. \cite{alvioli}. It clearly appears that, while the dotted line, 
obtained including only the effect of NN correlations in the quasi elastic 
channel fails to explain the measured cross section at low energy loss, the 
inclusion of FSI and inelastic channels brings theory and experiment into 
agreement over the whole $\omega$ range.

\begin{figure}[h]
\begin{center}
\epsfig{figure=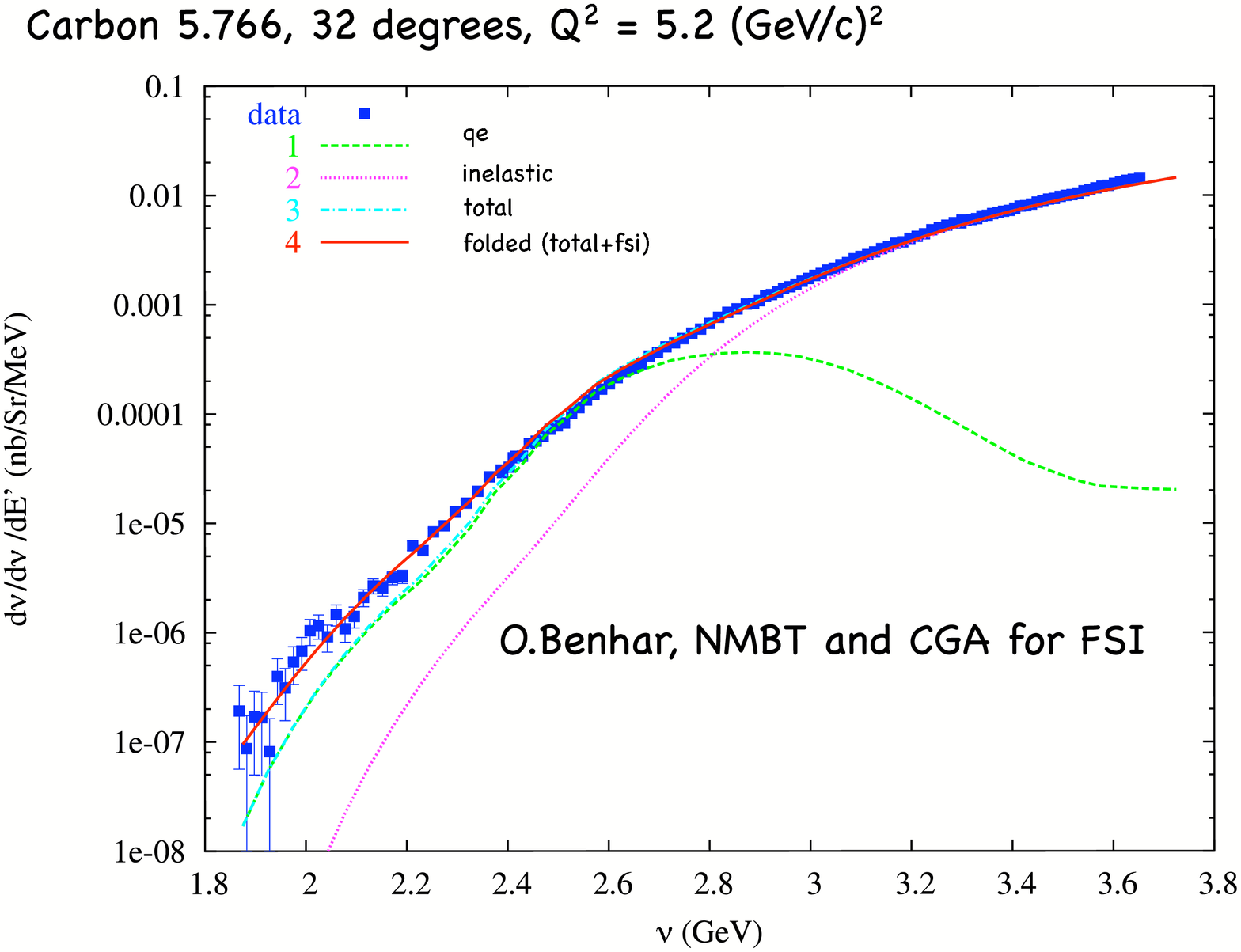,angle=270,width=9.5cm}
\end{center}
\caption{Comparison between the measured cross section of the process $^{12}C(e,e')X$ 
at beam energy $E = 5.8$ GeV and electron scattering angle 32$^\circ$ \cite{nadia} 
and the results of theoretical calculations based on the approach 
of Ref. \cite{benhar91}, represented by the solid line.
The labels of the bottom axis refer to the energy transfer. (After Ref. \cite{day07}). \label{jlab07} }
\end{figure}

The remarkable agreement between data and the results of theoretical 
calculations including {\em both} NN correlations and FSI is also illustrated in 
Fig. \ref{jlab07} \cite{day07}, showing the Carbon cross section at 
beam energy $E = 5.8$ GeV and scattering angle 32$^\circ$, measured 
at JLab \cite{nadia}. 

The results of Figs. \ref{claudio} and \ref{jlab07} strongly suggest that
a quantitative understanding of FSI is required to unambiguously
identify correlation effects in the region of low energy loss.

A procedure aimed at extracting information on NN correlations, i.e. on the 
high momentum components of the nuclear wave function, from the inclusive 
cross section at $x>1$ has been proposed in Ref. \cite{egiyan03,egiyan06}, whose 
authors argue that the appearance of a plateau in the ratio of the cross 
sections corresponding to different targets is due to the fact that 
the nuclear momentum distributions have a similar behavior at large momenta, 
and essentially differ only by an overall factor. Hence, according to 
Ref. \cite{egiyan03,egiyan06} the plateau is a signature of the cancellation of
FSI effects in the ratio.
\begin{figure}[h]
\begin{center}
\includegraphics[width=18.pc]{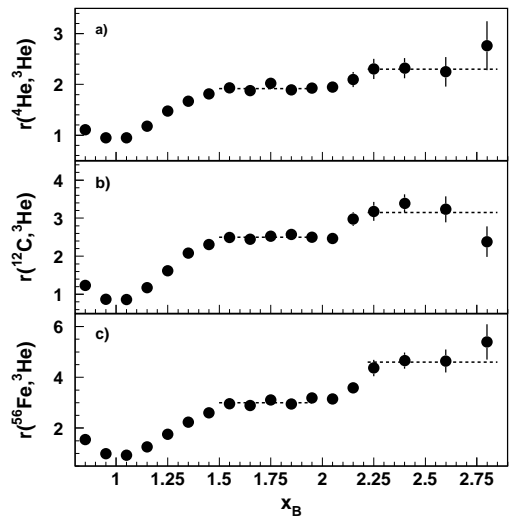}
\end{center}
\caption{Cross section ratios of (a) $^4$He, (b) $^{12}$C and (c) $^{56}$Fe 
to $^3$He as a function of the Bjorken scaling variable for $Q^2>1.4$ GeV$^2$.  
(After Ref. \cite{egiyan06}). \label{ratios} }
\end{figure}

As an example, Fig. \ref{ratios} shows the ratios
\begin{equation}
R_A = \frac{3}{A} \ \frac{d\sigma(e + A \rightarrow e^\prime + X)}
{d\sigma( e + ^3\!He \rightarrow e^\prime + X)},
\label{ratio}
\end{equation}
for $A$=4, 12 and 56, corresponging to $^4$He, $^{12}$C and $^{56}$Fe,
in the range $1<x<2.8$.
 It is of course tempting to interpret the strength near $x=2$
($x=3$) as originating from scattering off correlated two(three)-nucleon systems, 
with mass 2$m$ (3$m$). This interpretation, however,
ignores the fact that the data exhibit clear scaling in the variable
$y$ \cite{sick80}, hereby proving that
the electron scatters off constituents with nucleonic mass and nucleonic
form factor.
The interpretation of cross section ratios as ratios of
correlation strengths is also hindered by the fact that $x$,
unlike $y$, is not simply related to the momentum carried by the struck
nucleon \cite{liuti93}.

The assumption of cancellation of FSI effects, undelrlying the analysis proposed 
in Ref. \cite{egiyan03,egiyan06}, is still controversial. Establishing its validity
will require a systematic study of the ratios within the approaches 
which have proved succesful in explaining the measured cross sections.

\section{Electron-induced two-nucleon knockout}

The interest in the experimental investigation of two-nucleon emission 
processes was triggered, at the end of the 1980s, by the prospect of the upcoming
new generation of continuous beam electron accelerators \cite{benhar90}. 
In the following years a number of measurements of  
$(e,e^\prime NN)$ cross sections have been carried out at 
 NIKHEF-K~\cite{OnAl,Gerco,Ronald} and MAMI~\cite{ros00}.

Ideally, the two-nucleon knockout reaction can be regarded as the cleanest source of
experimental information on NN correlations, as it may give access to the 
two-nucleon spectral function \cite{benhar00}
\begin{equation}
\PkkE=\sum_n \vert \langle n|a_{{\bf k}_1}a_{{\bf k}_2}|
0 \rangle \vert ^2
\delta (E-E_n+E_0) \ , 
\end{equation}
yielding the probability of removing two nucleons of momenta ${\bf k}_1$ and
 ${\bf k}_2$ from the target ground state $|0\rangle$, leaving the residual 
$(A-2)$-nucleon system with excitation energy $E$.

However, it was soon realized that the role of correlations may be obscured by
the presence of competing mechanisms, such as FSI, and that 
extracting information on the spectral function from the data requires
i) a careful choice of the kinematical setup and ii) the development of 
consistent theoretical models including all the relevant effects.

The comparison between the $^{16}$O$(e,e'pp)$ cross sections measured at 
NIKHEF-K and theoretical
calculations~\cite{Ry97,PRC57-pp} has shown that 
correlation effects are dominant in the transition to the ground
state of $^{14}$C when the two protons are emitted back-to-back 
with small total momentum. The large cross section measured in this 
kinematical considtions is in fact regarded as a clean signature
of correlations.

Accurate theoretical calculations must include a consistent description
of both the two-nucleon overlap functions, containing the information
on correlations between the pair of knocked out nucleons in the initial state, 
 and the nuclear electromagnetic current. The effects of FSI of the 
outgoing nucleons, with one another and with the recoiling nucleus, must be also 
taken into acount.

The results of recent work carried out by the Pavia Group \cite{cm1,cm2} 
suggest that the 
requirement that single particle bound and scattering states be 
orthogonal may play a significant role. The orthogonalization 
procedure developed in Refs. \cite{cm1,cm2} takes into account the 
spurious contributions associated with the center of mass motion, and is 
therefore suitable for application to targets such as oxygen.

The relevance of the treatment of orthogonalization turns out to 
depend on the specific process and kinematical setup. 
 While in many instances the effect of the spurious contributions 
is small, it is very large in the so called super-parallel kinematics, 
in which the momenta of the two knocked out nucleons are parallel and anti-parallel 
to the momentum transfer ${\bf q}$. This kinematical 
 setup has been adopted in the measurements of the $^{16}$O(e,e$'$pp)$^{14}$C \cite{ros00}
and $^{16}$O(e,e$'$pn)$^{14}$N  \cite{Duncan} cross sections carried out at 
MAMI. 

\begin{figure}[h]
\begin{center}
\includegraphics[width=38.pc]{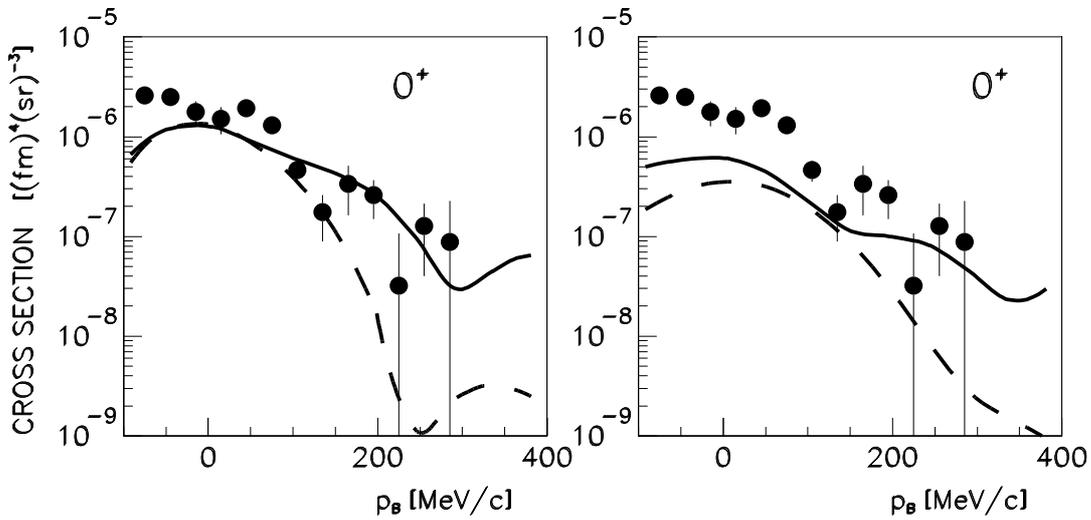}
\end{center}
\vspace*{-.1in}
\caption{Differential cross section of
the reaction $^{16}$O(e,e$'$pp)$^{14}$C in super-parallel kinematics
at beam energy $E=855$ MeV, energy transfer $\omega=215$ MeV and 
momentum transfer $|{\bf q}|=316$ MeV, as a function of the 
recoil momentum ${\bf p}_{\mathrm{B}}$. The solid lines have been 
obtained including the full FSI effect, while the results represented 
by the dashed lines do not take into account the interactions between
the two ejected nucleons. The left and right panels correspond to the
different othrogonalization procedures described in the text. The data 
is taken form Ref. \cite{ros00}.
(After Ref. \cite{giusti}). \label{eeNN} }
\end{figure}

As an example, Fig. \ref{eeNN} shows the differential cross section of
the reaction $^{16}$O(e,e$'$pp)$^{14}$C to the $0^+$ ground state in 
super-parallel kinematics. The data have been taken at incident electron energy 
$E=855$ MeV, energy transfer $\omega=215$ MeV and momentum transfer 
$|{\bf q}|=316$ MeV \cite{ros00}.
Different values of the recoil momentum ${\bf p}_{\mathrm{B}}$ correspond to 
different kinetic energies of the two outoging nucleons. Positive (negative)
values are associated with ${\bf p}_{\mathrm{B}}$ parallel (anti-parallel) 
to ${\bf q}$. 

The results of the orthogonalized approach with and without
removal of the the spuriosity are displayed in the left and
right panels, respectively. The solid lines have been 
obtained including the full FSI effect, while the results represented 
by the dashed lines do not take into account the interactions between 
the two ejected nucleons. It clearly appears that FSI effects are large and
a correct treatment of the center of mass motion is needed to a 
bring theory and data into agreement at ${\bf p}_{\mathrm{B}}>0$.

\section{Neutrino-nucleus scattering}

As pointed out in the Introduction, experimental searches of neutrino 
oscillations exploit neutrino-nucleus interactions
to detect the beam particles, whose properties are unknown.
The use of nuclear targets as detectors,
 while allowing for a substantial increase of the event rate,
entails non trivial problems, since data analysis requires a quantitative
understanding of the neutrino-nucleus interactions.
In view of the present experimental accuracy,
the treatment of nuclear effect is in fact regarded as one of the
main sources of systematic uncertainty. 

The description of nuclear dynamics is even more critical to neutrino
experiments aimed at obtaining {\em nucleon} properties from {\em nuclear} 
cross sections \cite{K2K,BOONE}.

Starting in 2001, a series of Workshops on ``Neutrino-Nucleus Interactions in 
the Few GeV Region'' (NUINT) has been devoted to the discussion of nuclear 
effects in neutrino interactions (see Ref.\cite{NUINT07} and References therein). 
The NUINT Workshops are mainly aimed at establishing a connection between 
the communities of electron- and neutrino-nucleus scattering, with the purpose
of generalizing the approaches successfully employed in the analysis of
electron-nucleus scattering to the case of neutrino scattering.
The simulation codes currently employed by many oscillation experiments are 
based on the Relativistic Fermi Gas Model (RFGM). According to this model, the 
nucleon momentum distribution $n({\bf k})$ is flat up to the 
Fermi momentum $k_F$, and vanishes at $|{\bf k}|>k_F$, while the
removal energy is fixed to a constant value $\epsilon$.

\begin{figure}[h]
\begin{center}
\includegraphics[width=20.pc]{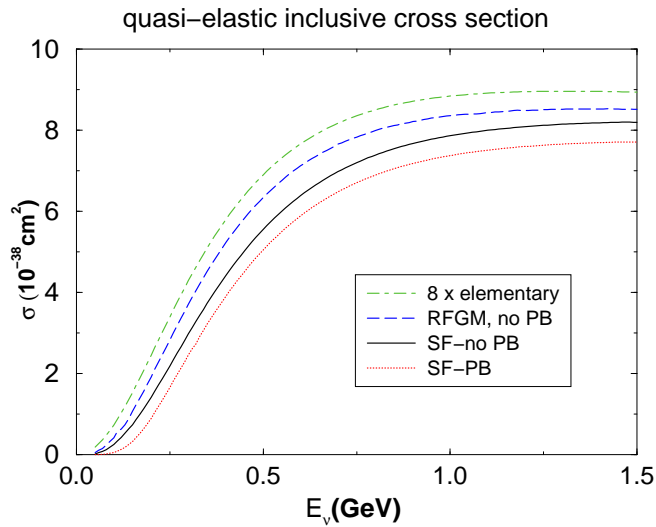}
\end{center}
\caption{Total quasi-elastic cross section of the process
$\nu_e + ^{16}O \rightarrow e^- +  X$.
The dot-dash line represents eight times the elementary cross section; the dashed
line is the result of the RFGM with Fermi momentum
$k_F = 225$ MeV and binding energy $\epsilon=25$ MeV; the
dotted and solid lines have been obtained using the spectral function of
Ref. \cite{benhar94}, with and without inclusion of Pauli blocking, respectively.
(After Ref. \cite{benhar07}). \label{nu_QE1} }
\end{figure}

\begin{figure}[h]
\hspace*{.1in}
\includegraphics[width=18pc]{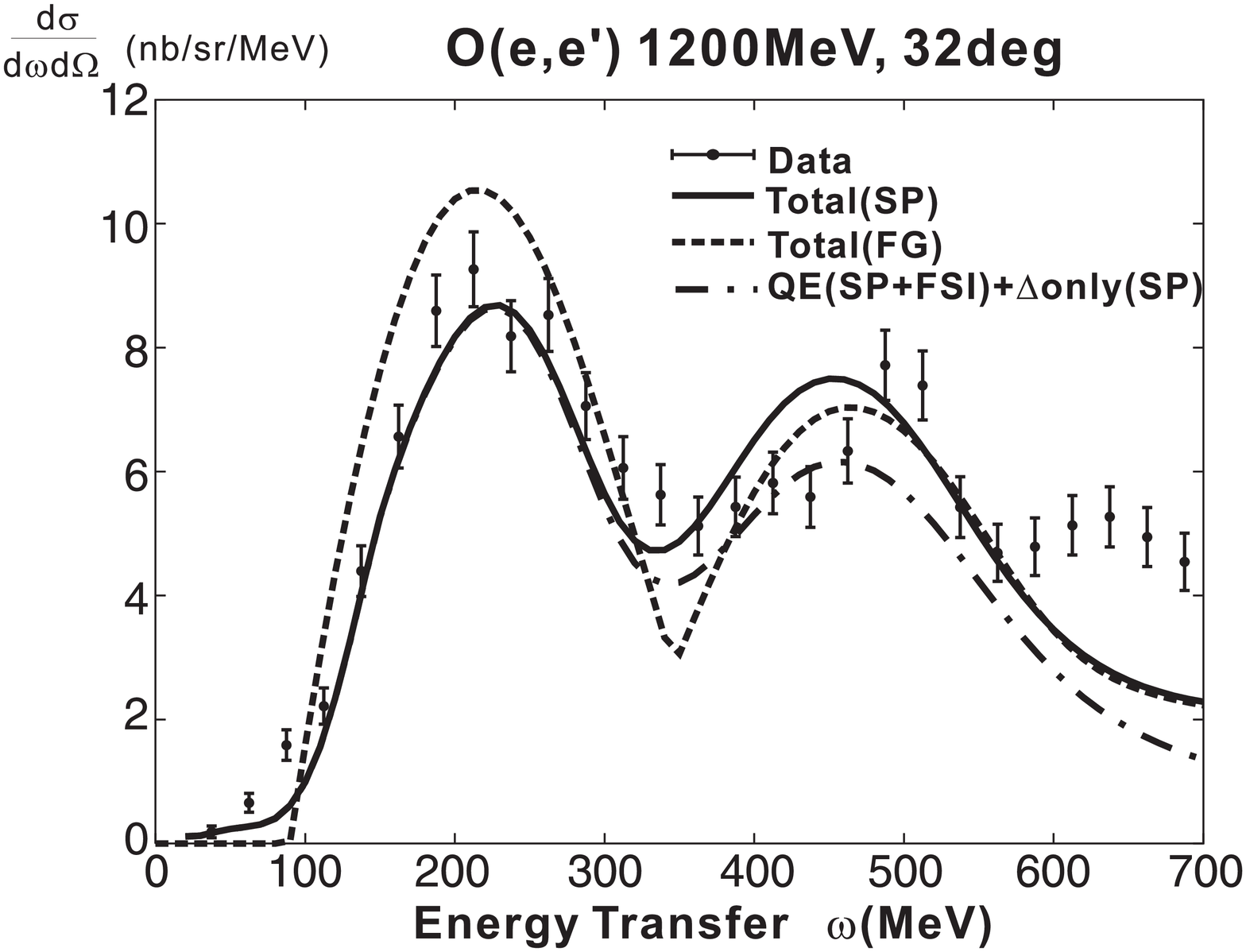}
\includegraphics[width=18pc]{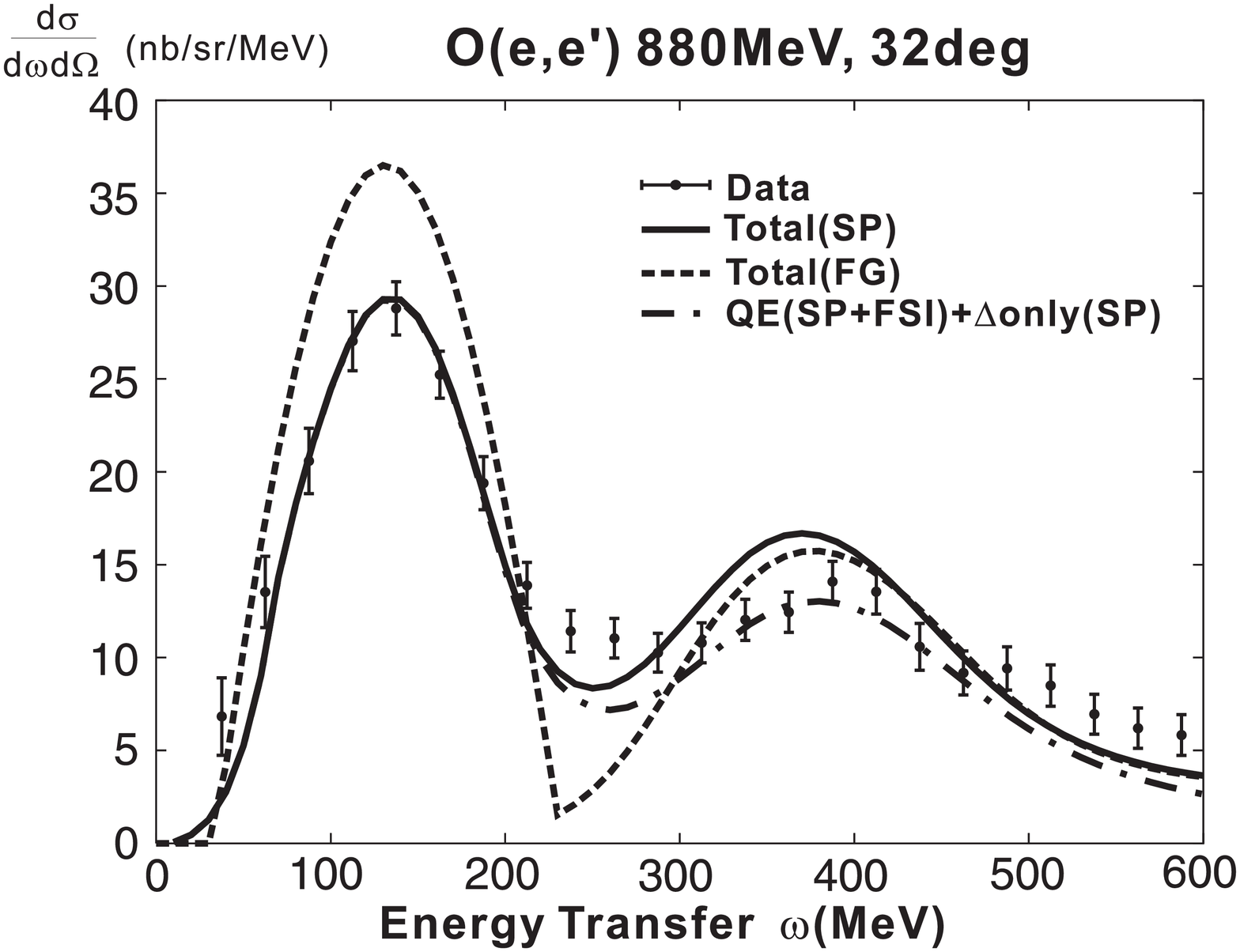}
\caption{Comparison between the measured cross section of the process
$^{16}O(e,e')X$ at electron scattering angle $\theta = 32^\circ$ \cite{lnf} 
 and the results
of theoretical calculations including the effects of both short range
correlations and FSI, represented by the solid lines. The Left and right
panels correspond to electron beam energy $E = 1.2$ GeV and $880$ MeV, 
respectively. For comparison, the results of the RFGM are also shown by 
dashed lines (After Ref. \cite{nakamura07}). \label{naka} }
\end{figure}

In order to gauge the magnitude of the nuclear effects under discussion, 
and the need of improving upon the RFGM description, 
Fig. \ref{nu_QE1} shows the energy dependence of the quasi elastic contribution 
to the total cross section of the process $\nu_e + ^{16}O \rightarrow  e^- + X$ 
computed using different approximations. The dot-dash line represents the 
result obtained describing oxygen as a 
collection of noninteracting stationary nucleons, while the dashed and 
solid line have been obtained from the RFGM and Eq.({\ref{dsigma:PWIA}), 
with the spectral function of Ref. \cite{benhar94}, respectively. 
It is apparent that replacing the RFGM with the approach based on
a realistic spectral function leads to a sizable suppression of the total
cross section. Comparison between the dot-dash line and the dotted one, 
obtained from a generalization of Eq.(\ref{dsigma:PWIA}) designed to include 
the effect of Pauli blocking \cite{benhar05}, shows that 
 the overall change due to nuclear effect is $\sim$~20~\%.

Note that FSI between the nucleon produced at the elementary weak interaction 
vertex and the spectator particles have not been taken into account, as they 
{\em do not} contribute to the total cross section.

\subsection{Validation of nuclear models through comparison to electron scattering 
data}

The accuracy of the models of nuclear effects to be used in 
the analysis of neutrino oscillation experiments can be tested by comparing 
theoretical results to electron scattering data. 

The relevant kinematical domain can be readily identified considering
the quantum mechanical phase difference developed by two neutrino
mass eigenstates over a distance $L$
\beq
\Delta \varphi_{jk} = (E_k - E_j) L \approx \frac{\Delta m^2_{jk}}{2E_\nu} L \ ,
\eeq
where $\Delta m^2_{12} \approx 6.9 \times 10^{-5} \ {\rm eV}$ and
$\Delta m^2_{23} \approx 2.5 \times 10^{-3} \ {\rm eV}$.
Knowing $\Delta m^2_{23}$, the energy $E_\nu$ that maximizes the 
oscillation can be obtained from 
\beq
E_\nu = 0.60 \ \left[ \frac{\Delta m^2_{23}}{3 \times 10^{-3}
\ {\rm eV^2}} \right] \ \left[
 \frac{L}{150 \ {\rm Km}} \right] \ {\rm GeV} \ ,
\eeq
showing that for long baseline experiments $E_\nu$ ranges between several 
hundreds MeV to $\sim 1.5$ GeV. In this kinematical region quasielastic scattering 
and pion production, mainly trough $\Delta$ excitation, are known to provide
the dominant contributions to the cross section \cite{lls}.

As an example, Fig. \ref{naka} shows a comparison between the cross section 
of the process $e + ^{16}O \rightarrow e^\prime + X$ at beam energy $E = 1.2$ GeV 
and $880$ MeV and electron scattering angle $\theta = 32^\circ$, measured at Frascati
\cite{lnf}}, and the results of the theoretical calculations of Nakamura {\em et al.} 
\cite{nakamura07}. The solid lines correspond to the full calculation,   
carried out within the approach based on a realistic spectral function, while 
the dashed lines have been obtained within the RFGM. 

The peaks corresponding 
to quasi elastic scattering and $\Delta$ production are both clearly visible 
in the data. While the former is reproduced  
 with an accuracy of $\sim$~10~\%, more sizable 
discrepancies between theory and data occur at $\omega$ above pion production 
threshold. The authors of Ref. \cite{benhar06} argued that they may be
ascribed to the lack of accurate parametrizations of the neutron structure
functions in the $\Delta$ production region at the low values of $Q^2$ 
corresponding to the data of Ref. \cite{lnf} $(Q^2 \sim 0.2$ GeV$^2$).

Independent of the description of the electron-nucleon vertex, however, 
the results of Fig. \ref{naka} clearly show that replacing the RFGM with
the approach of Ref. \cite{nakamura07} leads to a much better 
overall agreement between theory and data.

\subsection{Superscaling analysis}
\label{superscaling}

Scaling is observed in a variety of scattering
processes involving many-body systems \cite{west}. For example, at large
momentum transfer the response of liquid helium measured
by inclusive scattering of thermal neutrons, which in general depends upon
{\it both} ${\bf q}$ and the energy transfer $\omega$, exhibits a striking scaling
behavior, i.e. it becomes a function of the single variable
$y = (M/|{\bf q}|)(\omega - |{\bf q}|^2/2M)$, $M$ being the mass of the
helium atom \cite{he}. Scaling in a similar
variable occurs in inclusive electron-nucleus scattering
at $|{\bf q}| \gsim $ 500 MeV and electron energy loss $\omega < Q^2/2M$
 \cite{sick80}.
Another most celebrated example is scaling of the deep inelastic proton
structure functions, measured by lepton scattering at large $Q^2$,
in the Bjorken variable $x$ \cite{bj}.

Being a consequence of the   kinematics of the underlying electron-nucleon 
scattering process, scaling provides a strong handle on the reaction mechanism. 
Furthermore, the observation of scaling violations reveals that the dynamics 
go beyond the IA picture.

\begin{figure}[h]
\hspace*{.1in}
\includegraphics[width=18pc]{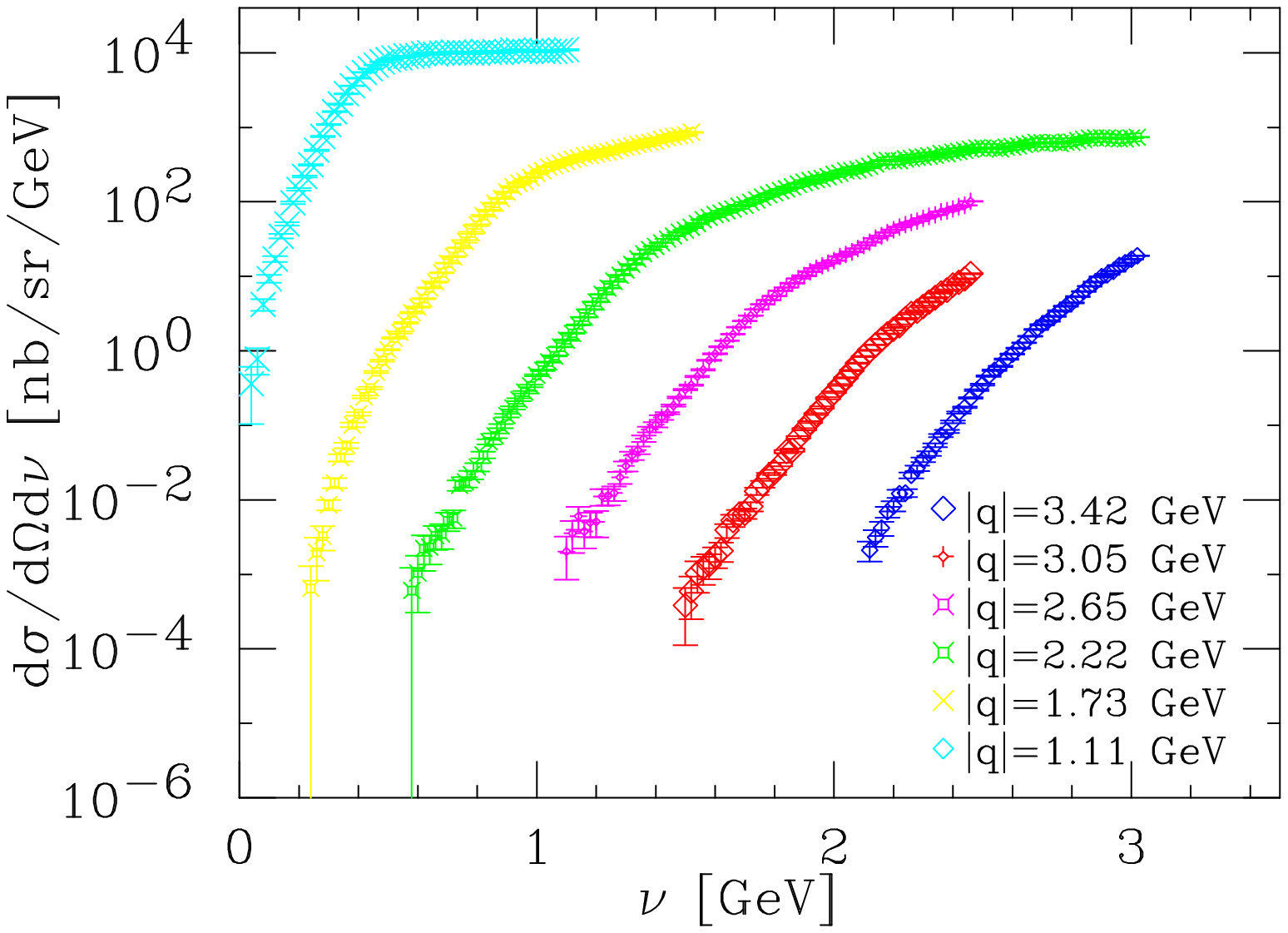}
\includegraphics[width=18pc]{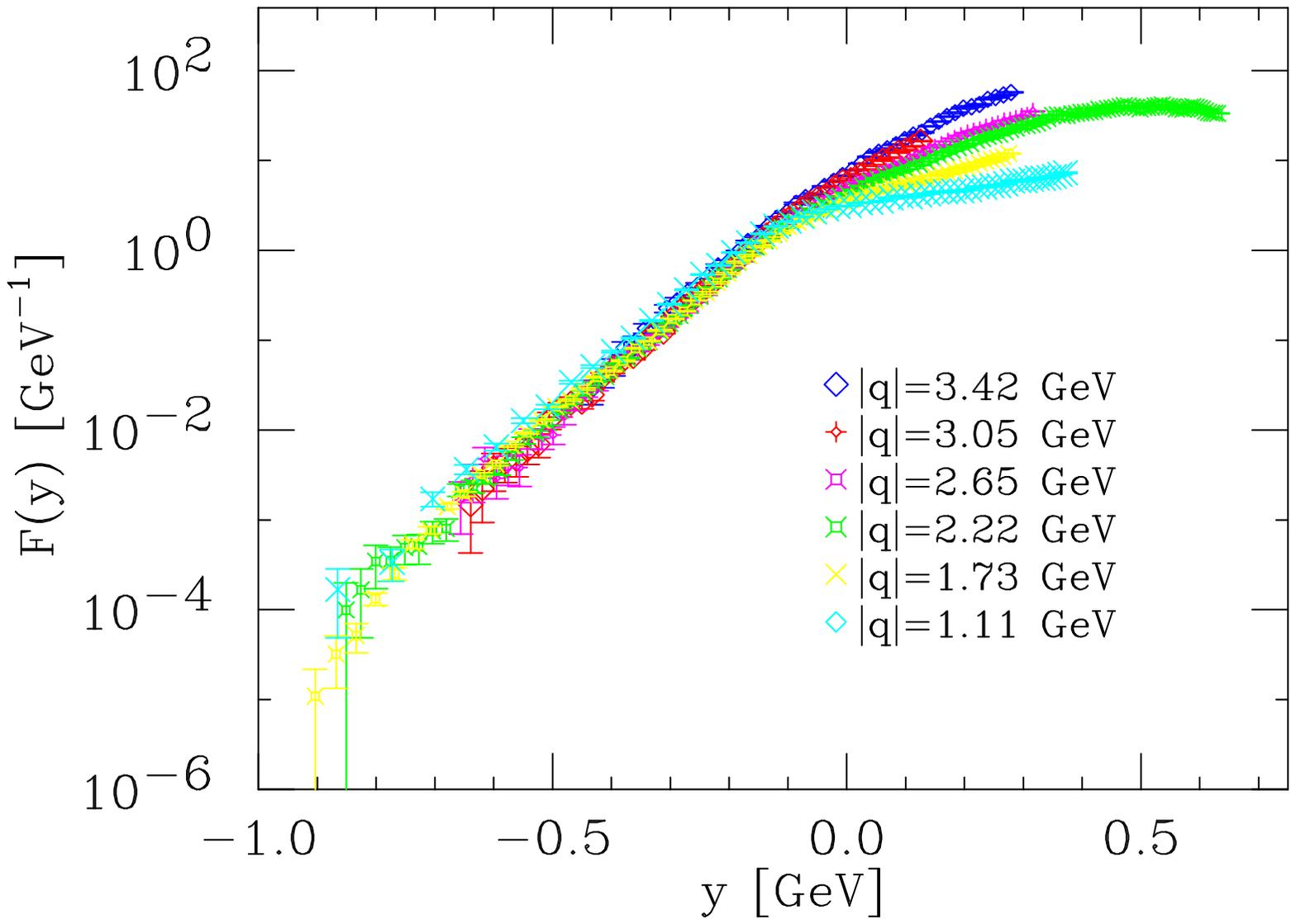}
\caption{Left panel: Inclusive cross sections of the process 
$^{56}Fe(e,e')X$ measured at JLab \cite{arrington}. The different curves
correspond to different values of the momentum transfer at the quasi free
peak. Right panel: $y$-scaling functions obtained from the cross sections 
shown in the left panel.\label{yscaling} }
\end{figure}

Figure \ref{yscaling} provides an illustration of $y$-scaling in electron 
nucleus scattering. It clearly appears that the inclusive cross sections
off Iron at momentum transfer ranging between $\sim$1 GeV and $\sim$3.5 GeV, 
spanning more than seven orders of magnitude, when plotted as a function of the 
scaling variable $y$ collapse to a single curve at $y<0$. The scaling 
violations observed in the region of positive $y$, corresponding to large 
electron energy loss, have to be ascribed to the occurrence of inelastic 
electron-nucleon scattering. The results of Fig. \ref{yscaling} suggest that 
the occurrence of $y$-scaling can be exploited to predict the inclusive 
cross section at any value of $|{\bf q}|$ in the kinematical regime in which 
IA is applicable

The scaling analysis has been recently pushed one step
further \cite{donnelly99}.
 Motivated by the Fermi gas model, in which all momentum distributions
only differ by an overall scale factor, the Fermi momentum, the authors
of Ref. \cite{donnelly99} have investigated whether the scaling functions of 
different nuclei can be related to one other, by adjusting one overall scale factor.
As it turned out, this appears to be possible for nuclei with $A \geq 12$. 

\begin{figure}[h]
\begin{center}
\includegraphics[width=19pc]{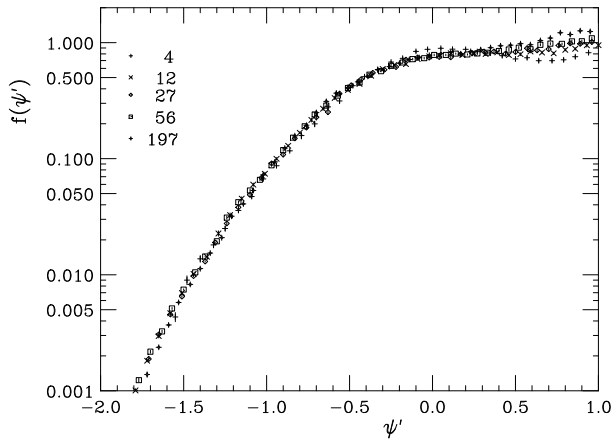}
\end{center}
\caption{Scaling function of nuclei with $A$ in the range $4-197$ at fixed 
kinematics, corresponding to $|{\bf q}|\sim $1 GeV, plotted 
as a function of the scaling variable $\psi^\prime$ (After Ref. \cite{benhar08}). 
\label{psiscaling} }
\end{figure}

Figure~\ref{psiscaling} shows an example of the scaling function $f(\psi^\prime)$, 
obtained from the analysis of the data of Ref. \cite{arrington},  
plotted as a function of the new scaling variable $\psi^\prime$, corresponding
to $y$ scaled by a Fermi momentum.

Scaling as a function of the nuclear mass number, called scaling of second kind, 
or {\em superscaling}, appears to be realized better than $y$-scaling, 
which is broken by the non quasi elastic contributions to the cross section. 
At the same kinematics, these contributions are in fact not too dissimilar for 
different nuclei. As the momentum used to scale $y$ is a slowly varying and 
smooth function of $A$, superscaling is very useful to interpolate data and 
predict the cross sections off nuclei not experimentally investigated. 

Inclusive electron scattering and charged current neutrino
scattering are closely related, the underlying nuclear physics being the
same. Hence, it has been suggested that superscaling may be exploited to 
accurately predict the cross sections of neutrino induced reactions relevant 
to the ongoing experimental activity (see, e.g. Ref. \cite{bai0}).

\begin{figure}[h]
\begin{center}
\epsfig{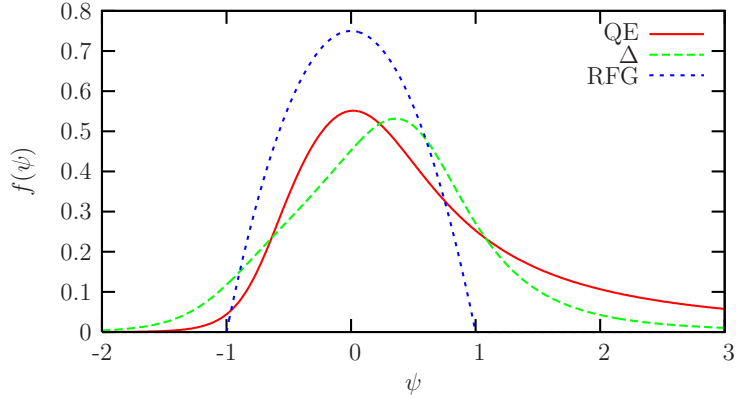}
\end{center}
\caption{Superscaling function resulting from the analysis of electron
scattering data. The solid and long-dashed lines correspond to the
quasi elastic and the $\Delta$-production contributions, respectively.
For comparison, the short-dashed line shows the results of the RFGM.
(After Ref. \cite{bai1})\label{scal1}}
\end{figure}

As shown in Fig. \ref{naka}, however, in the kinematical region relevant to 
long baseline oscillation experiments both quasi elastic scattering and
$\Delta$-production contribute to the cross section. 
The extension of the superscaling approach to the $\Delta$-production 
region, carried out in Ref. \cite{amaro05}, 
is based on the observation that the cross section exhibits a pronounced
peak at $\omega = (|{\bf q}|^2 + m_\Delta^2)^{1/2}$, $m_\Delta$ being the 
$\Delta$ mass, whose width is determined by the nuclear Fermi momentum.
Figure \ref{scal1} shows the quasielastic and $\Delta$-production 
superscaling functions obtained from the analysis of electron
scattering data. 

\begin{figure}[h]
\begin{center}
\epsfig{figure=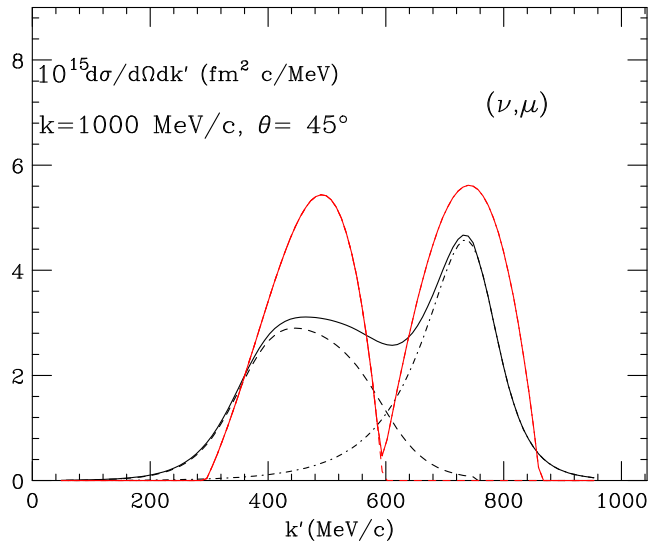,angle=0,width=8.5cm}
\end{center}
\caption{Charged current interaction neutrino-nucleus cross section 
corresponding to beam energy $E_\nu =$ 1 GeV and muon
emission angle $\theta =$ 45$^\circ$, plotted as a function of
the muon energy. The dot-dash and dashed lines correspond
to the quasi elastic and $\Delta$-production contributions, respectively, while 
the solid line represents the total cross section. For comparison, the 
result of the RFGM is also shown by a solid line (After Ref. \cite{bai1}) 
\label{scal2} }
\end{figure}

As an example of the application of the supercaling analysis to 
neutrino-nucleus scattering, Fig. \ref{scal2} shows the charged current 
interaction cross section corresponding to beam energy $E_\nu =$ 1 GeV and muon 
emission angle $\theta =$ 45$^\circ$.

The comparison with the predictions of the RFGM, also
displayed in both Figs. \ref{scal1} and \ref{scal2}, clealry indicates that 
nuclear dynamics plays a critical role and should not be described using 
oversimplified approaches. Although the idea of superscaling has been inspired
 by the RFGM, the scaling functions turn out to be quite different from 
the ones obtained from this model. As pointed out in Ref. \cite{martini}, 
 this is a clear indication that some physical effects not included in the 
RFGM, while not affecting the appearance of scaling, are in fact non negligible. 

\begin{figure}[h]
\begin{center}
\epsfig{figure=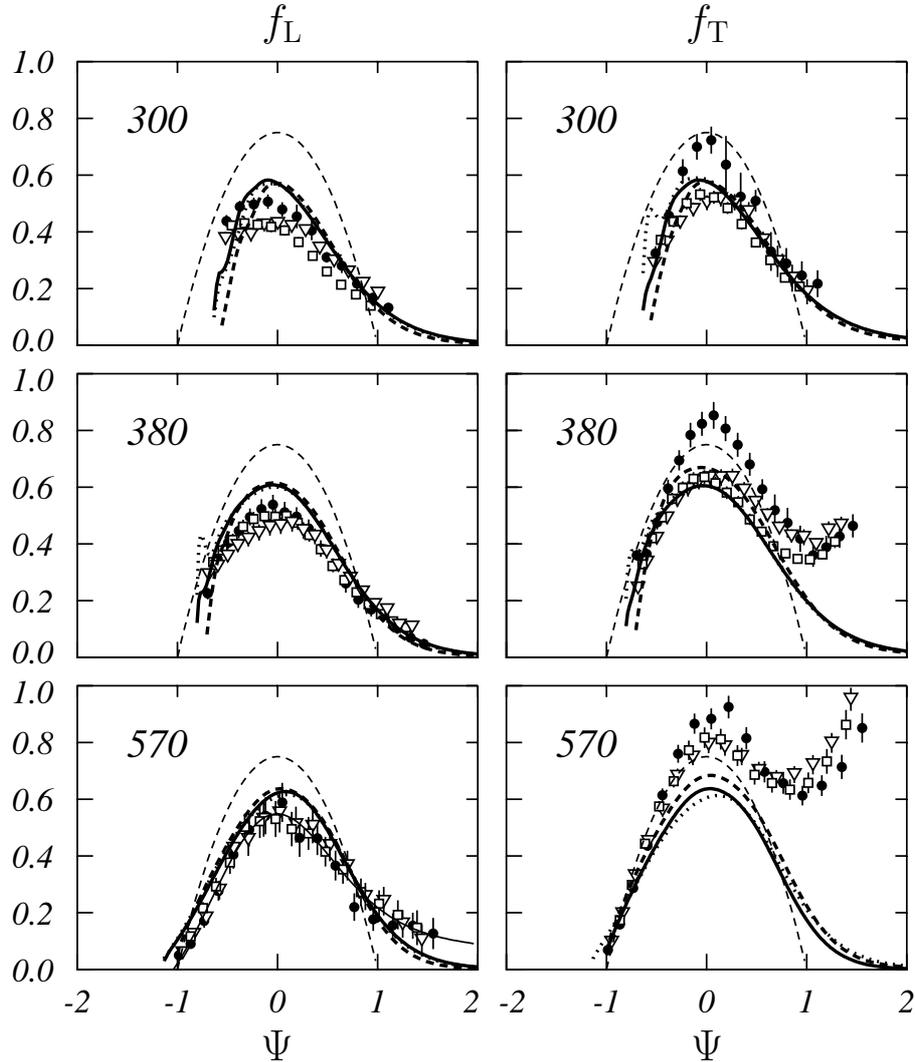,angle=0,width=12.5cm}
\end{center}
\caption{Longitudinal, $f_{\rm L}$, and transverse,
  $f_{\rm T}$, scaling functions obtained from the data of
  Ref. \cite{jou96}. The panels are labelled according to 
  the value of the momentum transfer in MeV. Full
  circles, white squares and white triangles refer to $^{12}$C, $^{40}$Ca and 
  $^{56}$Fe , respectively. The thin black line in the $f_{\rm
  L}$ panel at 570 MeV represents the empirical scaling function obtained
  by fitting the data, while the thick lines show the results 
  including effects beyond the RFGM.
  The full, dotted and dashed lines correspond to 
  $^{12}$C, $^{16}$O, and $^{40}$Ca, respectively. For comparison, the thin 
  dashed lines show the RFGM scaling 
  functions, scaled by a Fermi momentum. (After Ref. \cite{martini}). 
\label{scal3} }
\end{figure}

Figure \ref{scal3} shows the results of calculations including a number of
effects neglected in the RFGM: the finite size of the system, its collective 
excitations, NN correlations, FSI and meson exchange currents \cite{martini}. 
It is apparent that the inclusion of these effects produces sizable 
modifications of the scaling functions, with respect to the RFGM results. 
These modifications remarkably improve the
agreement with the experimental scaling functions. On the other hand,
they do not heavily affect the scaling behavior of the functions 
$f_{\rm L}$ and $f_{\rm T}$. 

\subsection{Deuteron breakup by low energy neutrinos}

Ulike the long baseline neutrino oscillation experiments discussed in the previous 
Section, the experiments aimed at detecting solar neutrino oscillations are sensitive to 
nuclear interactions of low energy neutrinos.

The search for oscillations of solar neutrinos carried out by the SNO 
collaboration \cite{SNO} is based on the determination of the yield 
of the neutral and charged current deuteron disintegration processes
\be
\nonumber
\nu_x+d&\longrightarrow&\nu^\prime_x+n+p \ , \\
\nonumber
\nu_e+d&\longrightarrow&e^-+p+p \ ,  
\ee
at neutrino energy $E_\nu$ up to  $\sim 20$ MeV.
Extracting the information on the solar neutrino flux 
from the data requires the knowledge of the cross sections
of the neutrino- and antineutrino-deuteron breakup cross section.

Accurate theoretical calculations of these processes have been 
carried out using currents derived from elementary hadron amplitudes, 
extracted in the tree approximation from the chiral Lagrangians,
and nuclear wave functions generated from realistic nuclear
potentials \cite{MRT1}. 

In principle, the alternative approach based on effective field theories 
should be regarded as more fundamental. However, it involves parameters 
that cannot be determined from processes involving elementary particles. 

Recently, Mosconi {\em et al} have analized the uncertainties associated
with the calculations carried out within the two different approaches. 
Their work was aimed at assessing the model dependence of the theoretical
results, that contributes to the systematic error of the experiment. 

\begin{figure}[h]
\begin{center}
\epsfig{figure=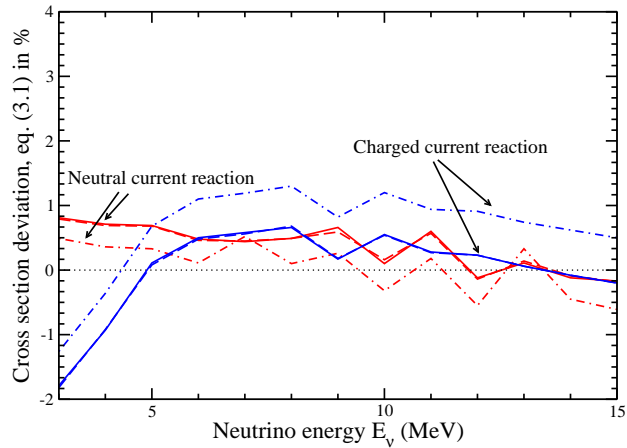,angle=270,width=9.5cm}
\end{center}
\caption{Energy dependence of the quantity $\delta^a_i$, defined by
Eq.(\ref{def:delta}), providing a measure of the differences between 
the deuteron breakup cross sections obtained from potential models and 
effective field
theories. (After Ref. \cite{mosconi07}). \label{mosconi} }
\end{figure}

Figure \ref{mosconi} shows the energy dependence of the quantity
\beq
\delta^a_i =  \ 1-\frac{\sigma^a_{EFT}}{\sigma_{pot,i}} \ ,
\label{def:delta}
\eeq
yielding a measure of the difference between the cross sections 
obtained from potential models, $\sigma_{pot,i}$, and effective field 
theories, $\sigma^a_{EFT}$.

Although the uncertainty turns out to be quite small 
for both the neutral current and charged current breakup, it is comparable 
to the full effect of the two-body currents. Based on this observation, the 
authors of Ref. \cite{mosconi07} conclude that the accurate determination of the 
effect of the two-body currents is still an open issue.

\section{Summary and prospects}

The italian theoretical nuclear physics community, keeping up its 
long-standing tradition, is carrying out first 
class research in the field of electron-nucleus scattering, 
working in many instances in close connection with experimental 
collaborations.

After more than three decades, a significant effort is being devoted to 
the study of the effects of NN correlations, which still elude a precise
experimental determination. 
At the Workshop on Short Range Structure of Nuclei at 12 GeV, 
held at Jlab in the fall of 2007, it was suggested that, in order  
to understand the interplay between correlation and 
FSI contributions to the inclusive cross section at $x\gg1$, theorists
should agree on a well defined {\em homework} problem, 
to be used as a test case to compare the results of various approaches.
Such a comparison is definitely much needed, as in the past different 
formalisms have been mostly used to obtain different observables.

While inclusive processes are likely to help identify correlation effects, 
the $(e,e^\prime p)$ cross section appears to be better suited to carry out a 
direct measurement of the tails of the nucleon spectral function at high 
momentum and high removal energy \cite{rohe04}. More exclusive 
experiments, measuring the double coincidence $(e,e^\prime NN)$ cross sections, 
may even provide direct access to the internal dynamics of pair a correlated 
nucleons. 

In spite of the many difficulties associated with these
experiments, as well as with the consistent and realistic 
theoretical description of the two-nucleon emission process, we are finally
approaching the level of development required to make significant comparisons 
between theory and data. This will help to pin down the role
of the competing reaction mechanisms, as well as the kinematical setups in which 
NN correlations dominate.

The italian groups have been quick to realize that the available 
theoretical and experimental information on electron-nucleus 
scattering can be very useful in the analysis of neutrino oscillation 
experiments. Their contributions, clearly visible in the Proceedings
of the Workshops of the NUINT series (see 
Ref. \cite{NUINT07} and References therein) 
are likely to have a large influence in shaping the field.

The approaches based on the direct calculation of the netrino-nucleus cross 
section using realistic spectral functions, as well as the superscaling analysis of 
electron-scattering data, appear to describe the scattering process much 
better than the RFGM, currently employed for the data analysis of
most long baseline neutrino experiments. 

In spite of their success, however, it has to be realized that the potential
for exploiting more advanced nuclear models in 
the analysis of oscillation experiments, in which the nucleus is seen as a
detector rather than a target, largely depends upon the possibilty 
of implementing the formalism in Monte Carlo simulations. This is likely to 
be one of the main focuses of the research activity in the years to come.

\section*{Acknowledgements} 
I wish to express my gratitude to the organizers of the XII Workshop on 
 Theoretical Nuclear Physics in Italy, for inviting me to a very interesting 
and productive meeting. A number of illuminating discussions with the participants 
in the meeting of the INFN Theory Collaboration PI31, held at ECT*, Trento,
on March 27-28, 2008, are also gratefully acknowledged.

\end{document}